\documentclass[fleqn,twoside]{article}

\usepackage{amssymb}

\usepackage{espcrc2}

\usepackage{graphicx}
\usepackage[figuresright]{rotating}

\newcommand{\AmS}{{\protect\the\textfont2
  A\kern-.1667em\lower.5ex\hbox{M}\kern-.125emS}}

\title{
  Neutrino Factory R\&D
  \thanks{
    Talk presented at the High Intensity Frontier Workshop, La Biodola,
    Isola d'Elba, Italy, $28^{\rm th}$ May -- $1^{st}$ June 2005.
  }
}

\author{
  K.Long 
  \address[Imperial]{
    High Energy Physics Group,
    Department of Physics,
    Imperial College London,
    Exhibition Road,
    London SW7 2BZ.
  }
}

\begin{document}

\begin{abstract}

Elegant experiments are being carried out, or are in preparation, to
improve the precision with which the solar and atmospheric
neutrino-oscillation parameters are known, and to attempt to make a
first measurement of the small mixing angle $\theta_{13}$. 
The compelling case for the development of an accelerator-based
neutrino source to serve the programme of precision measurements
of neutrino oscillations and sensitive searches for leptonic-CP
violation that is required to follow these experiments is briefly
reviewed.
The Neutrino Factory, an intense high-energy neutrino source based on
a stored muon beam, is widely believed to yield a precision and
sensitivity superior to other proposed second-generation facilities.
The alternatives are identified and the case for a critical comparison
of the performance of the various options is presented.
Highlights of the exciting international R\&D programmes which are
designed to demonstrate the feasibility of the required techniques are
then reviewed. 
The steps that the international community is taking to produce,
by the end of the decade, a full conceptual design for the facility
are described.
The ambition of the Neutrino Factory community is to demonstrate the
feasibility of a cost-effective design such that, should forthcoming
measurements show that it is required, the facility could be brought
into operation in the second half of the next decade.
\vspace{1pc}

\end{abstract}

\maketitle

\section{Introduction}
\label{Sect:Intro}

\begin{figure}[htb]
  \includegraphics[width=1.0\columnwidth]{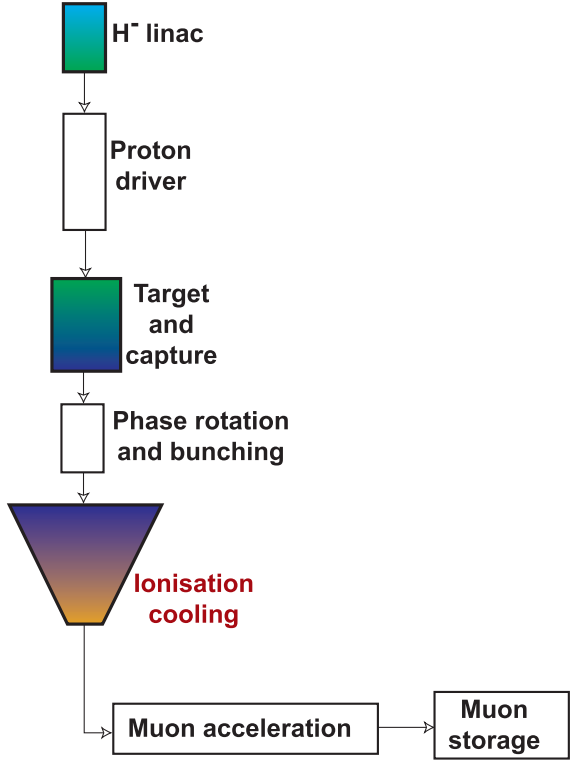}
  \caption{
    Schematic drawing showing the major sub-systems in the Neutrino
    Factory accelerator complex.
          }
  \label{Fig:NFSchematic}
\end{figure}

Beams of high-energy electron- and muon-neutrinos will be produced
from intense stored muon beams at the Neutrino Factory
\cite{Ref:Geer}.  
A schematic diagram of the main sub-systems of the accelerator
facility is shown in figure \ref{Fig:NFSchematic}. 
The process of generating the stored muon beam starts with the
bombardment of a suitable target with a high-power pulsed proton
beam of moderate energy ($\sim 5 - 15$~GeV). 
Pions and kaons produced in the target are captured and allowed to
decay to produce muons; the muons must be accelerated
rapidly to $\sim 20 - 50$~GeV before being injected into the
storage ring.
The muon beam initially occupies a very large phase space, making it
necessary to develop fast, affordable, large-aperture acceleration
systems and/or a phase-space reduction (cooling) technique that is rapid
when compared to the muon lifetime.
The feasibility of such a Neutrino Factory has been addressed in a
number of studies \cite{Ref:FSI,Ref:FSII,Ref:FSIIa,Ref:DSEU,Ref:DSJp}.
These studies defined the programme of R\&D required to establish
technological solutions for each of the key accelerator systems.
The main purpose of this contribution to the proceedings of the High
Intensity Frontier Workshop is to review the status of these R\&D
programmes (section \ref{Sect:NF}). 

In making the case that the Neutrino Factory is the tool of choice for
the era of precision measurements of neutrino oscillations
and high-sensitivity searches for leptonic-CP violation,
it is necessary to review the Neutrino Factory physics reach in the
context of the present generation of experiments and the performance
of other proposed second-generation neutrino facilities (such as
second-generation super-conventional beams
\cite{Ref:MWWS,Ref:FNALPDWS} or beta-beams \cite{Ref:BetaBeam}). 
Furthermore, the importance of the study of the neutrino is such that
it is essential to bring into operation the best possible
second-generation programme at around the time the present-generation
experiments are being completed.  
The experimental programme planned for the next five years is
discussed in section \ref{Sect:Motivation} and used to identify the
timescale on which it would be desirable to make the case for the
Neutrino Factory.

\section{Motivation}
\label{Sect:Motivation}

\subsection{Phenomenology of neutrino oscillations}
\label{SubSect:Pheno}

The phenomenon of neutrino oscillations is readily described by
extending the Standard Model (SM) to include three neutrino mass
eigenstates; $\nu_1$, $\nu_2$, and $\nu_3$ with masses $m_1, m_2$ and
$m_3$ respectively \cite{Ref:NuOscPheno}. 
The flavour eigenstates, $\nu_e$, $\nu_{\mu}$, and $\nu_{\tau}$, are
obtained by rotating the mass eigenstates using the unitary matrix
$U$ which may be written:
\begin{equation}
  U = U_{23} U_{13} U_{12}
\end{equation}
where
\begin{eqnarray}
 U_{23} & = & \pmatrix{
               1 &  0      & 0                                     \cr
               0 &  c_{23} & s_{23}                                \cr
               0 & -s_{23} & c_{23},                               \cr
           }, \\
  U_{13} & = & \pmatrix{
                 c_{13}               & 0 & s_{13} e^{i \delta}    \cr
                 0                    & 1 & 0                      \cr
                -s_{13} e^{-i \delta} & 0 & c_{13}                 \cr
           }, \\
  U_{12} & = & \pmatrix{
                c_{12} & s_{12} & 0                                \cr
               -s_{12} & c_{12} & 0                                \cr
                0      & 0      & 1                                \cr
           },
\end{eqnarray}
the cosines and sines of the three mixing angles $\theta_{12}$,
$\theta_{13}$, and $\theta_{23}$ are denoted $c_{12}$ etc., and
$\delta$ is a phase parameter.
Measurements of neutrino-oscillation probabilities can not be
used to determine the absolute neutrino-mass scale but, since the
probabilities depend on the mass-squared differences, 
$\Delta m^2_{23} = m^2_3 - m^2_2$ and 
$\Delta m^2_{12} = m^2_2 - m^2_1$, neutrino oscillations can be used
to determine the mass hierarchy.
Electron neutrinos produced in the sun undergo elastic scattering with
electrons in the material of the sun.
This, the MSW effect \cite{Ref:NuOscPheno}, modifies the effective mass
that appears in the electron-neutrino oscillation probability and has
been used to determine the sign of $\Delta m^2_{12}$.
The sign of $\Delta m^2_{23}$ can be determined in oscillation
experiments for which the baseline is sufficiently long ($\gtrsim 1000
{\rm~km}$) and for which the neutrino energy is sufficiently high
($E_{\nu} \gtrsim 10 {\rm~GeV}$). 

Leptonic-CP violation will occur if $\delta \ne 0$ 
(and $\sin \theta_{13} \ne 0$).
Measurements of the difference between the oscillation probabilities
for neutrinos and anti-neutrinos can be used to determine $\delta$.
Such measurements require large data sets and appropriately chosen
baselines and neutrino energies. 
To obtain the required neutrino-interaction rate requires either a
very large detector or a very intense source (or both).

The challenge to the neutrino community is to measure all the mixing
angles of the MNS matrix as precisely as possible, to determine the
sign of $\Delta m^2_{23}$ and to measure precisely $\Delta m^2_{12}$
and $\Delta m^2_{23}$, and, by  measuring $\delta$, to discover
leptonic-CP violation if it occurs.
The fundamental importance of the search for leptonic-CP violation is
self-evident.
Precision measurements of the parameters that govern neutrino
oscillations are essential if a complete understanding of the nature
of the neutrino is to be obtained.
Such measurements will either establish the minimal model outlined 
above or, by establishing parameter sets inconsistent with it, point 
to the existence of entirely new phenomena; for example, the 
three-generation scenario would have to be abandoned should 
MiniBOONE \cite{Ref:MiniBOONE} confirm the presently unexplained LSND
result \cite{Ref:LSND1,Ref:LSND2,Ref:LSND3,Ref:LSND4}. 
The second-generation neutrino source must, therefore, be capable of
doing both.

\subsection{Context and timescales}
\label{SubSect:ContTime}

Data from the Sudbury Neutrino Observatory (SNO)
\cite{Ref:SNO1,Ref:SNO2} and KamLAND \cite{Ref:KamLAND1,Ref:KamLAND2}
experiments, together with data from Super-Kamiokande \cite{Ref:SK1} and
elsewhere have been used to determine $\theta_{12}$ with a precision
of around 10\% and $\Delta m_{12}^2$ with a precision of 10\% --
20\%. 
The parameters $\sin^2 \theta_{23}$ and $\Delta m_{23}^2$ have been
determined using atmospheric neutrino data from Super-Kamiokande
\cite{Ref:SK2} and verified using an accelerator-based neutrino source
by the K2K experiment \cite{Ref:K2K}. 
With five to seven years of running, the MINOS long-baseline
experiment \cite{Ref:MINOS1,Ref:MINOS2}, which has begun to take data,
will determine $\theta_{23}$ and $\Delta m_{23}^2$ with a precision of
around 10\%. 
The two CNGS experiments OPERA \cite{Ref:OPERA} and ICARUS
\cite{Ref:ICARUS1,Ref:ICARUS2}, which are designed to observe
$\nu_\tau$ appearance and are scheduled to start data taking in 2008,
will verify aspects of the mixing formalism outlined above.  
Two first-generation super-beam experiments, T2K in Japan
\cite{Ref:T2K1,Ref:T2K2} and NO$\nu$A in the US \cite{Ref:NOVA}, are
being mounted with the objective of demonstrating that $\sin^2 2
\theta_{13}$ is greater than zero. 
The T2K experiment will start in 2009 and, after five
years of data taking, will be sensitive to $\sin^2 2 \theta_{13}$ down
to about 0.005 at 90\% C.L. NO$\nu$A, which has recently been granted
scientific approval by the FNAL PAC, will yield a comparable
sensitivity. 
Both T2K and NO$\nu$A will improve the determination of $\theta_{23}$
and $\Delta m_{23}^2$ to the level of a few percent after five years
of data taking.  
However, neither T2K (Phase I) nor NO$\nu$A will have the sensitivity
required to discover leptonic-CP violation or to deliver the precision
measurements of the parameters that are required for a full
understanding of neutrino oscillations.  

To take the study of neutrino oscillations further requires a
second-generation facility ready to begin operation in the second half
of the next decade. 
This facility must be capable of making high-precision measurements of
the mixing angles and mass-squared differences and of making searches
for leptonic-CP violation of great sensitivity. 
The precision of the measurements must be such that sensitive tests of
the consistency of the theoretical framework can be made. 
Three types of facility have been proposed to provide the neutrino
beams required to serve this second-generation programme.
The Neutrino Factory gives the best performance over almost all of the
parameter space and is believed to be the `facility of choice'. 
Second-generation super-conventional-beam experiments may be an
attractive option in certain scenarios. 
A beta-beam, in which electron neutrinos (or anti-neutrinos) are
produced from the decay of stored radioactive-ion beams, in
combination with a second-generation super-beam, may be competitive
with the Neutrino Factory \cite{Ref:NuFact}. 

Following the feasibility studies that were carried out at the turn of 
the century, an international programme of R\&D into the accelerator
complex has grown up, fostered in part by the `NuFact' (Neutrino
Factory, super-beam and beta-beam) workshop series which was initiated
in 1999.
The programme of hardware development, reviewed below, is now reaching
maturity.
To put in place the facility (or facilities) required to serve the
second-generation programme of precision measurement requires that a
conceptual design be prepared by the end of the decade together with
as broad a consensus as possible on the roadmap for its
implementation.
A step on this road was taken at NuFact05 with the launch of a
one-year international `scoping study' of a future Neutrino Factory
and super-beam facility \cite{Ref:ISS}.
The objectives of the scoping study are to \cite{Ref:ISSDoc}:
\begin{itemize}
  \item Evaluate the physics case for a second-generation super-beam,
        a beta-beam facility and the Neutrino Factory and to present a
        critical comparison of their performance; 
  \item Evaluate the various options for the accelerator complex with
        a view to defining a baseline set of parameters for the
        sub-systems that can be taken forward in a subsequent
        conceptual-design phase; and to 
  \item Evaluate the options for the neutrino detection systems with a
        view to defining a baseline set of detection systems to be
        taken forward in a subsequent conceptual-design phase. 
\end{itemize}
The conclusions of the scoping study will be presented at NuFact06 and
published in a written report in September 2006.

\section{Neutrino Factory R\&D}
\label{Sect:NF}

It is not possible in a short article such as this to do justice to
the Neutrino Factory R\&D programmes that are being carried out in
Europe, Japan and the US which, together, cover all aspects of the
facility. 
The following paragraphs therefore emphasise the key elements of the 
programme.

\subsection{The proton-driver front-end}
\label{SubSect:PDFE}

The Neutrino Factory proton driver is required to deliver $1-4$~MW of
proton-beam power at an energy of $5 - 15$~GeV in $\sim 1$~ns
bunches.
Machines of similar specification are required to drive a
next generation spallation-neutron source, a radioactive heavy-ion
facility, and to generate intense `super-conventional' neutrino beams.
Futhermore, high-power proton sources are required to serve
applications such as the transmutation of nuclear waste. 
Such CW sources share many of the technological challenges presented
by high-power pulsed proton beams.

The activation of the accelerator elements through the loss of a
fraction of the primary beam power is the principal issue for the
development of a high-power proton driver.
To keep the activation within acceptable limits requires that the
beam-loss rate should be no more than 1~W/m \cite{Ref:Findlay}.
To achieve this challenging specification requires that the beam
quality at injection be exceptionally good.
Several programmes aimed at developing the technologies
required to produce such high quality beams are underway
\cite{Ref:FETS}. 
The programmes emphasise the front end of the accelerator, i.e. from
the ion source up to energies of a few MeV. 
For pulsed proton beams, the development of high-quality beam
choppers is of particular importance.
`Choppers' are designed to remove unwanted bunches from the beam
with 100\% efficiency and are required if low-loss injection into
accumulator or compressor rings or clean on-off transitions are to be
achieved.

Such parallel developments are a strength as they allow the sharing of
expertise and information and give confidence that the front-end of
the Neutrino Factory accelerator complex will be developed on an
appropriate timescale. 

\subsection{Target and capture}
\label{SubSect:TrgtCptr}

Efficient pion production may be achieved by bombarding a rod-like
high-$Z$ material with the primary proton beam.
For solid targets, fatigue caused by beam-induced thermal shock is the
principal issue that must be addressed in the design of the target
\cite{Ref:TrgtShock}. 
To reduce the effect of shock damage to solid targets may require that
the target be replaced every beam pulse.
Several solid-target schemes have been proposed
\cite{Ref:TrgtSchemes}.
A free-flowing liquid-mercury-jet target is a conceptually simple
alternative \cite{Ref:TrgtSchemes}. 
Shock-induced processes cause the break-up of the jet,
therefore the jet velocity must be chosen such that a new volume of
liquid mercury is exposed to the beam every pulse.

Two schemes have been proposed by which the particles produced in the
target may be captured.
The first uses high-field solenoid magnets to capture both positive
and negative particles at the same time \cite{Ref:FSII}.
The second calls for a magnetic horn to focus either positive or
negative particles into the subsequent transport and decay sections
\cite{Ref:DSEU}. 
The horn scheme has the advantage that the focussing element closest
to the target itself is relatively simple.
The advantage of the solenoid scheme is that an efficiency gain of a
factor of two can be achieved if the downstream accelerator complex
is designed to manipulate and store $\mu^+$ and $\mu^-$ simultaneously
\cite{Ref:FSIIa}.
In each case, significant engineering work needs to be carried out to
ensure that the target station can be operated safely.

Particle production in the target has been studied
\cite{Ref:Brookes}.
Though particle-production models give significantly different rates
and spectra, the results indicate that a proton driver with an energy
in the range $\sim 5 - 15$~GeV is likely to be suitable.
In order to optimise the target and capture system, it will be
important to bench-mark the various simulation codes against measured
particle distributions.
Two experiments, HARP \cite{Ref:HARP} at CERN and MIPP
\cite{Ref:MIPP} at FNAL have been (or are being) carried out to
measure these distributions. 
The HARP experiment has recently presented results for
forward-particle production \cite{Ref:HARPLA}. 
The large-angle data, which is expected to be finalised soon, will be
important in tuning the particle-production models.

\subsubsection{Characterisation of materials}
\label{SubSubSect:CharMat}

The development of the conceptual design for the Neutrino Factory
target station rests on an understanding of the properties of the
various proposed materials under extreme conditions.
Irraditation studies of solid targets are being carried out at BNL and
at CERN \cite{Ref:TrgtSchemes}.
These studies include the investigation of the degree to which the
bombarded material can be annealed by baking at high temperature.
The intensity, repetition rate, and beam time available for these
studies are insuffucient to simulate target exposures comparable to
long-term (several months to a year) use at the Neutrino Factory.
The UK Neutrino Factory collaboration is therefore developing a
technique in which a high-current pulse is used to generate,  in a
sample of tantalum wire, energy densities comparable to those expected
in the Neutrino Factory target \cite{Ref:TrgtIPulse}.
The current-pulse technique will be used to mount a `life-time' test.
The numerical techniques needed to extrapolate these measurements to
the Neutrino Factory target using  LS-DYNA \cite{Ref:LSDYNA} are being 
developed in parallel.

Measurements of the effect of intense proton-beam pulses on liquid
mercury have been carried out at BNL, and studies of the development
of mercury jets both with and without magnetic field have been carried
out at Grenoble and at BNL respectively \cite{Ref:OldLHg}.
For liquid-jet targets, the energy deposited by the beam can be
sufficient to cause voids to be created in the body of the jet by the
shock-induced transient pressure waves.
This process, referred to as cavitation, is being studied
experimentally at CERN using a high-power laser impinging on a jet of
water \cite{Ref:Lettry}.
Numerical studies of the passage of particle beams through mercury-jet
targets have been developed and now give a good description of the
measured behaviour \cite{Ref:JetSim}.

\subsubsection{The liquid-mercury-jet target}
\label{SubSubSect:LHg}

The liquid-metal option for a pion-production target capable of
operating with a multi-MW pulsed proton beam at a Neutrino Factory
will be tested in the MERIT experiment \cite{Ref:MERIT}. 
MERIT, which has recently been given scientific approval at CERN and
will be carried out by an international collaboration, will expose a
mercury jet of 1~cm diameter and flowing at 20~m/s in a 15~T
solenoidal magnetic field to an intense proton beam from the CERN PS. 
MERIT is scheduled to begin to take data in 2007.

\begin{figure}[htb]
  \includegraphics[width=1.0\columnwidth]{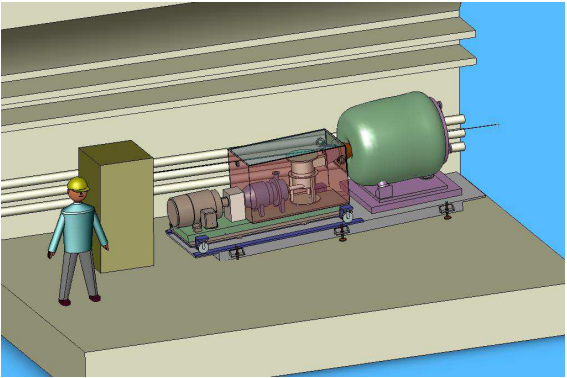}
  \caption{
    Schematic diagram of the MERIT experiment at the neutron
    time-of-flight (nToF) facility at CERN \cite{Ref:MERIT}.
          }
  \label{Fig:MERIT}
\end{figure}

A schematic diagram of the experiment is shown in figure
\ref{Fig:MERIT}. 
The proton beam from the PS is horizontal and enters the experiment
from the right. 
The experiment is tilted so that the angle between the proton beam and
the mercury-jet axis is 100~mrad.
Liquid nitrogen will be used to cool the copper coils of the solenoid
magnet to 80~K.
The magnet, pulsed with a 5~MVA power supply, will deliver a 15~T
field for a duration of 1~s.
The mercury jet will be injected into the 15~cm diameter warm bore of
the magnet and the beam-target interaction will be recorded through
viewing ports by high-speed cameras via fibre-optic cables.

Short bunch trains containing between one and four bunches of 
$5 - 7 \times 10^{12}$ protons will be extracted from the PS.
If all four bunches are filled, a total of $28 \times 10^{12}$ protons
will impinge on the target within a 2~$\mu$s spill giving a peak
energy deposition of 180~J/g. 
By varying the pattern of filled proton bunches the experiment will
also be able to study the effect of cavitation.

\subsection{Ionisation cooling}
\label{SubSect:IonCool}

The muon beam that emerges from the decay channel fills a large phase
space. 
For example, in US Study II the transverse emittance at the exit of
the decay channel is 12~mm \cite{Ref:FSII}.
The spread of the muons in the longitudinal phase space is also very
large ($\sim 60$~mm in Study II). 
Efficient, cost effective, acceleration of the muon beam requires that
the phase space be modified.
The phase-rotation and bunching systems that follow the decay channel
are required to produce a beam with an energy spread of $\sim 60$~MeV
which is appropriately bunched to match the subsequent cooling
sections. 

Each of the five Neutrino Factory conceptual design studies have
considered the benefit of reducing the emittance of the muon beam
(cooling) before injecting it into the acceleration and storage
systems.
There are two principal motivations for this: to increase the number
of muons inside the acceptance of the downstream accelerators; and to
keep the cost of the muon acceleration system to a minimum.

At the end of the decay channel, the muons have a momentum of roughly
200~MeV/c.
The time-dilated lifetime of the muon is short ($\sim 4.7 \mu {\rm s}$)
making it essential that cooling and acceleration take place as
rapidly as possible. 
Ionisation cooling, a process in which the muon beam is caused to pass
through an alternating series of liquid-hydrogen absorbers and
accelerating RF-cavities, is the technique by which it is proposed to
cool the muon beam prior to acceleration. 
Various `gain factors' have been defined to quantify the gain in
performance due to the cooling channel (see table
\ref{Tab:CoolFact}). 
Systems that give gain factors of between 2 and 10 have been devised.
Since a factor of $\Gamma$ gain in stored muon-beam intensity implies
a reduction, by a factor $\Gamma$, in the running time required to
achieve a particular total neutrino flux, and a decrease in emitance
of the muon beam entering the acceleration section is likely to lead
to significantly lower costs for muon acceleration, it will be
important to make a careful optimisation, for performance and cost, of
the cooling and acceleration systems.
The engineering demonstration of the ionisation-cooling technique will
be carried out by the international Muon Ionisation Cooling Experiment
(MICE) collaboration \cite{Ref:MICE}.
The MICE experiment, which has been approved, will take place at the
Rutherford Appleton Laboratory (RAL), using muons produced by the ISIS
800~MeV proton synchrotron.
The status of the experiment is reviewed in the paragraphs that
follow.
\begin{table*}[htb]
  \caption{
    Survey of the gain afforded using ionisation cooling in a
    number of conceptual design studies of the Neutrino Factory.
           }
  \label{Tab:CoolFact}
  \begin{tabular}{ccccl}
    \hline
Design & Number of      & Gain factor   & Cooling & Comment                \\
       & cooling cells  & per cell (\%) &         &                        \\
    \hline
Study II \cite{Ref:FSII}   & 26 &  6 & 7 & Increase in phase-space density \\
                           &    &    &   & in acceptance of downstream     \\
                           &    &    &   & accelerator.                    \\
    \hline
Study IIa \cite{Ref:FSIIa} & 26 &  2 & 2 & Increase in number of muons     \\
                           &    &    &   & in acceptance of subsequent     \\
                           &    &    &   & muon acceleration section.      \\
    \hline
CERN \cite{Ref:DSEU}       & 36 & 10 & 7 & Increase in muon yield at       \\
                           &    &    &   & 2~GeV over optimised Neutrino   \\
                           &    &    &   & Factory without cooling.        \\
    \hline
NuFact-J \cite{Ref:DSJp}   & -- & $1.5 - 2$ 
                                    & -- 
                                        & Acceleration based on FFAGs.    \\
                           &    &    &   & Performance improvement when    \\
                           &    &    &   & absorber is included in FFAG    \\
                           &    &    &   & ring giving 6D cooling effect.  \\
    \hline
  \end{tabular}
\end{table*}

\subsubsection{The international Muon Ionisation Cooling Experiment}
\label{SubSubSect:MICE}

\begin{figure*}[htb]
  \includegraphics[width=2.\columnwidth]{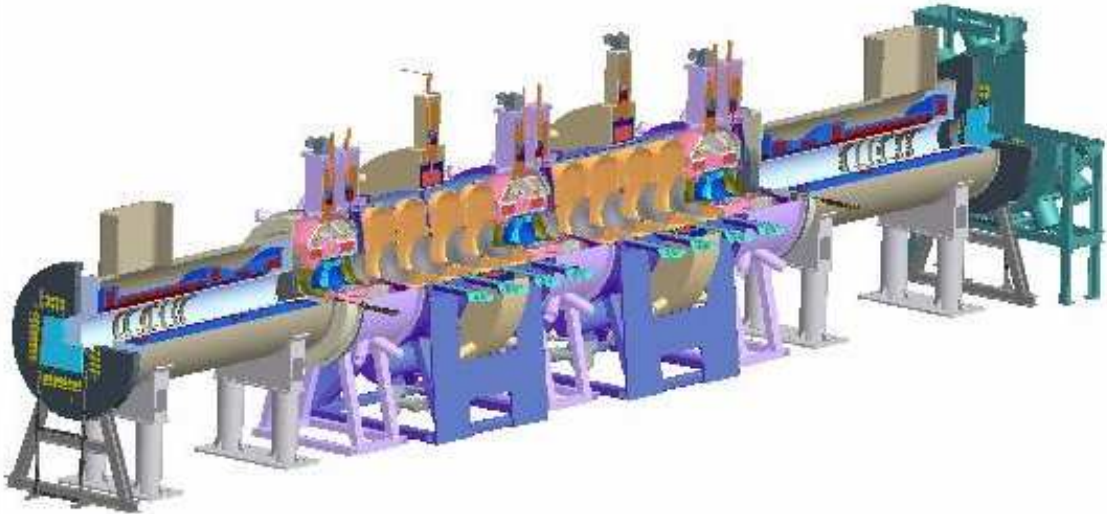}
  \caption{
    Drawing of the MICE experiment \cite{Ref:MICE}. 
    The beam enters the experiment from the bottom left-hand corner.
    The beam first passes through one of the scintillator hodoscopes
    that form the time-of-flight system.
    After passing through the upstream spectrometer, the beam passes
    through three absorber/focus-coil modules and two cavity/coupling-
    coil modules before it passes through the downstream spectrometer
    and a second time-of-flight hodoscope, the downstream Cherenkov
    counter and is stopped in the electromagnetic calorimeter.
    The beamline and upstream instrumentation are not shown.
          }
  \label{Fig:MICE}
\end{figure*}

The principal components of the MICE experiment are shown in figure
\ref{Fig:MICE}.
Two, functionally equivalent, spectrometers are placed upstream and
downstream of a single lattice cell of the Study II cooling channel.
In the Study II design approximately $10^{14}~\mu / s$ pass through
the channel.
The lateral dimensions of the beam are such that
space-charge forces can be ignored making it possible to run MICE
as a single particle experiment in which the Neutrino Factory bunch is
reconstructed offline using an ensemble of particles recorded in the
experiment. 
At the nominal input emittance of $\epsilon_{\rm in} = 6 \pi~{\rm mm}$
a cooling effect ($\epsilon_{\rm out} / \epsilon_{\rm in} - 1$, where 
$\epsilon_{\rm out}$ is the output emittance) of $\sim 10\%$ is
expected. 
The cooling effect will be measured with a precision of 1\% (i.e. 
$(\epsilon_{\rm out} / \epsilon_{\rm in} - 1)$ will be measured with
an absolute precision of 0.1\%).

The MICE cooling channel consists of three absorber/focus-coil (AFC)
modules and two accelerating-cavity/coupling-coil (RFCC) modules.
The AFC modules each contain a 21~l liquid-hydrogen absorber inside a
pair of superconducting coils that bring the beam to a focus in the
centre of the absorber.
Liquid hydrogen is the most efficient ionisation-cooling material
because it has a large specific ionisation and a comparatively large
radiation length.
Safe operation of the system in the presense of liquid hydrogen leads
to significant engineering constraints.
The AFC modules and the hydrogen system each have both active and
passive safety systems. 
The hydrogen will be stored in the form of metal hydride when the
absorber is emptied.
A vigorous R\&D programme is underway to demostrate the safe
operation of the hydrogen system. 
The super-conducting coils and the liquid-hydrogen vessel itself are
refrigerated using closed-cycle `cryo-coolers' \cite{Ref:CryoCooler}.

The RFCC module must restore the energylost by the muons as they pass
through the absorber.
The coupling coil, a short, large diameter solenoid, provides the
magnetic field that transports the muons through the module.
The acceleration is produced by four 201~MHz copper cavities which
produce a gradient of 8~MV/m.
To produce the required field gradient, the cavities must be
electrically closed, yet, to preseve the cooling effect, the amount of
material through which the beam passes must be minimised.
Thin berylium windows have been developed for this purpose. 
The degree of emission from the cavity surfaces is significantly
enhanced by the Lorentz force produced by an intense magnetic field
\cite{Ref:Norem}. 
While reducing the field emission in a Neutrino Factory cooling
channel, in which the cavities must operate at 16~MV/m, is a
challenging problem, it has been estimated that for operation in MICE,
the emission can be kept within acceptable bounds.

The muon beam that enters the experiment may contain a small pion
contamination.
The instrumentation upstream of the cooling channel is therefore
required to distinguish pions from muons and to measure the phase
space coordinates of the muons entering the channel.
Downstream of the cooling channel, the instrumentation is required to
identify electrons produced in the channel by muon decays and to
measure the muon phase-space coordinates.
The upstream particle identification will be performed using a
scintillator-based time-of-flight (TOF) system and a threshold
Cherenkov counter. 
The TOF system will also be used to trigger the experiment and to
determine the phase of the RF fields in the cavities as the muon
traverses the experiment.
The upstream and downstream spectrometers are each composed of a 4~T
superconducting solenoid instrumented with a scintillating-fibre
tracking device.
Downstream of the cooling channel a final TOF station, a Cherenkov
counter and a calorimeter are used to distinguish muons and electrons.

The MICE collaboration will take enough data to make the
uncertainty on the measured cooling effect systematics limited. 
It is therefore crucial that the systematic errors are understood in
detail. 
To do this, the experiment will be built up in stages.
A first measurement of cooling, using the two spectrometers and one
AFC module, is scheduled for 2008.
The first RFCC module and a second AFC module will then be installed
and the full MICE cooling channel will be assembled in 2009.

The MICE experiment will be mounted on ISIS at the CCLRC Rutherford
Appleton Laboratory.
The preparation of the MICE Muon Beam on ISIS, the MICE Hall and the
first phase of the MICE experiment are proceding to schedule.
The first data-taking period, in which the muon beam will be
characterised, the instrumentation calibrated and the relative
systematics of the two spectrometers will be measured, will begin in
April 2007. 

\subsection{Acceleration and storage}
\label{SubSect:AccelandStore}

The short lifetime of the muon leads to the requirement that the
acceleration be as rapid as possible.
Past studies have considered re-circulating linear accelerators in
various topologies.
More recently it has been proposed that fixed field alternating
gradient (FFAG) accelerators may offer advantages \cite{Ref:FFAG}.
Unlike conventional synchrotrons, the magnets within the FFAG are not
ramped with the consequence that the radius of the particle orbit
increases during acceleration.
The radial profile of the magnet pole-pieces is carefully designed to
give a field that varies with radius so as to produce the same
focussing effect for all momenta.
Several `scaling' FFAGs, in which the magnetic field scales with
radius, have been built in Japan \cite{Ref:FFAGJapan}.
The scaling FFAG programme is reaching maturity with the machine
proposed for the PRISM experiment \cite{Ref:PRISM}.
An alternative to the scaling FFAG is the `non-scaling' FFAG
\cite{Ref:NonScale}.  
In a non-scaling machine the magnets are standard quadrupoles or
combined-function dipoles.
However, the settings are carefully optimised so as to reduce the
radial displacement of the beam during acceleration.
An international collaboration (EMMA) is developing a proposal to
construct a proof-of-principle non-scaling FFAG \cite{Ref:EMMA}.

Progress has also been made in developing schemes for the storage
ring, the design of which is complicated by the need to serve two or
more detectors at different long base-lines \cite{Ref:Accel}.
To fully engineer the storage ring will require that the power
radiated from the muon beam in the form of decay electrons and
bremsstrahlung photons be dealt with using appropriate absorbers.

The experience gained in the construction of PRISM and EMMA and in the
design of the storage rings themselves will be important input to a
future Neutrino Factory design study in which the cooling,
acceleration and storage systems will have to be optimised together
for performance and cost. 

\section{Conclusions}
\label{Sect:Conclusions}

A programme of precision measurement of the properties of the neutrino
is important because the measurements may lead to the discovery of CP
violation in the lepton sector and of the physical principles that
explain the tiny neutrino masses and the very large neutrino mixing
angles can.
It is likely that these measurements will have a profound impact in
astro-physics and cosmology, well beyond the confines of particle
physics. 
The Neutrino Factory offers better sensitivity and precision than
other second generation facilities, and the accelerator systems
required are being developed by an energetic international community.
The time is therefore right for the Neutrino Factory community to take
the next bold step, to produce a conceptual design report by the end
of the decade.
If the community is successful in establishing the conceptual design
around the end of the decade and the results of the present generation
of experiments confirms that the Neutrino Factory is needed, then the
case to expedite the construction of the Neutrino Factory will be very
strong indeed.

\section*{Acknowledgements}

I would like to thank the organisers for giving me the opportunity
to present this review.
I gratefully acknowledge the help, advice, and support of my many 
colleagues within the international Neutrino Factory community who
have freely discussed their results with me and allowed me to use
their material.

\end{document}